\documentstyle [11pt]{article}
\newcommand{\beq}{\begin{equation}}
\newcommand{\eq}{\end{equation}}
\newcommand{\beqa}{\begin{eqnarray}}
\newcommand{\eqa}{\end{eqnarray}}

\begin{document}

\title{Two-body matrix elements of Pauli-projected excentric
single-particle orbits}
\author{K\'alm\'an Varga
\\
Institute  of Nuclear Research of the Hungarian Academy of
Sciences,
\\
P. O. Box 51, Debrecen, H--4001, Hungary}
\maketitle
\begin{abstract}
We present explicit analytical formulae for two-body matrix elements
between Pauli-projected single-particle orbits generated by gaussians
shifted from the centre of a nuclear core, which is populated by
nucleons occupying harmonic-oscillator orbits. Such shifted gaussian
orbits appear in the generator-coordinate model of two-cluster
systems and, as ingredients of generating functions, in the
resonating group model as well. A Pauli projection is to be
understood as a projection onto the subspace not occupied by the core
nucleons. This is an exact procedure to take into account the Pauli
principle in the limit in which the core is infinitely heavy, thus
the procedure to be presented is very useful in the microscopic
description of large-core plus small-cluster systems, such as
involved in $\alpha$-decay problems.
\end{abstract}

\vskip 24pt
\begin{center}
\S {\bf 1. Introduction}
\end{center}
\vskip 12pt

The more complex the nuclear system, the more complicated it is to
allow for the Pauli principle. In single-configuration models, like
the extreme shell model,
the Pauli principle is exactly allowed for by choosing the
single-particle (s.p.) orbits orthogonal to each other
and prescribing single population; hence the
interpretation of the Pauli principle as an exclusion principle.
For multiconfiguration one-centre models with a passive and massive
core, like the multiconfiguration shell model, the Pauli principle
between the core and valence particles is exactly observed
by choosing the valence orbits orthogonal to those occupied by the
core particles. However, for a multicentre system, such as a two-cluster
system, the antisymmetrization between the clusters
has to be carried through explicitly, and that implies tremendous
technical complications~\cite{hori}.

In this paper we consider the case when one of the clusters, the
``core", is substantially heavier than the other, ``the cluster".
In the limit, in which the core is completely passive and infinitely
heavy, the Pauli principle can again
be taken into account by an orthogonality
requirement, but it is not so easy to fulfil this requirement.
The subject of this paper is just to show a technique for the
treatment of the most critical aspect of such a model, that of the
calculation of the two-body potential matrix elements.

The wave function of the core $A$ is taken to be an
harmonic-oscillator (h.o.) shell model Slater determinant, and
that of the cluster $a$, which is assumed to contain at most four
nucleons, is restricted to be 0s h.o. orbits around a centre
displaced from the centre of the core. The two-centre shell model
underlying this picture forms a basis of the generator-coordinate
model (GCM) of the two-cluster system. The orthogonality
condition is imposed on the s.p. functions of cluster
$a$ by an operator that projects out the states filled in the core.

Though this model may look rather restrictive, it can serve as a
starting point for more sophisticated models. In particular, the
condition of the infinite mass of the core can be lifted by applying
a transformation that shifts the centre of mass (c.m.) of the
system from the centre of the core to the joint c.m.~\cite{hori}.
By further transformations the GCM can be replaced by a
resonating-group model~\cite{hori} or by one in which the relative
motion is represented by an h.o. basis~\cite{kami} or by a
single-centre basis spanned by gaussians of different
widths~\cite{kami}. Any of these transformations only involves the
generator coordinates (and not the physical ones), so they do not
interfere with antisymmetrization and can be performed on the matrix
elements. All of these transformations can be performed analytically,
but they should be performed before the angular-momentum projection.

The model with Pauli-projected s.p. orbits is useful especially for
the description of core+$\alpha$ systems. Since the relation of the
Pauli-projected orbits to the core orbits is the same as that of the
valence shell-model orbits, the Pauli-projected orbits can be
treated, to some extent, as shell-model orbits. In particular, the
configurations they can form are just like ordinary shell-model
configurations. The only difference is that they are not orthogonal
to the ordinary valence orbits, thus a diagonalization over a
non-orthogonal basis is required. Such cluster configurations have
been used in the description of $\alpha$
decay~\cite{Wild,Varga1,Varga2}.

To calculate the overlaps and s.p. matrix elements between such
Pauli-projected s.p. orbits is straightforward with the well-known
techniques~\cite{hori}. The calculation of the two-body
matrix elements is, however, not so simple. The two-body matrix
elements reduce to integrals of products of h.o. and shifted gaussian
functions. It seems to be natural to expand the gaussians in terms of
h.o. functions~\cite{Wild} and then use the conventional technique, but
this recipe leads to infinite summations, and the numerical
expenditure is enormous. The procedure we show avoids this,
by evaluating these integrals directly.

The plan of the paper is as follows. In \S 2 we
briefly summarize the essentials of the method of the Pauli-projected
s.p. orbits. In \S 3 we express the matrix element of an interaction
term as a sum of sixteen terms and derive a general formula for these
terms. In \S 4 we specialize this formula for
some terms, and in \S 5 we give a brief summary. The paper is concluded
with an Appendix containing a glossary of the formulae we used in the
derivation.

\vskip 24pt
\begin{center}
\S {\bf 2. Pauli-projected single-particle orbits}
\end{center}
\vskip 12pt

A general state of the $a+A$ two-cluster system can be written in the
form $\Psi\allowbreak={\cal A}\{\phi_{LM}({\bf r})\allowbreak\Phi_a(\xi_a)
\Phi_A(\xi_A)\}$, where $\Phi$ are antisymmetrized cluster internal
states, the function $\phi$ describes the relative motion of angular
momentum $LM$ and $\cal A$ is an intercluster antisymmetrizer
normalized in the pattern of ${\cal A}=N^{-1/2}\sum_{i=1}^N(-1)^{P_i}P_i$.
A GCM representation can be obtained by expanding
$\phi_{LM}({\bf r})$ in terms of angular-momentum projected gaussians
displaced with respect to the centre:
\begin{equation}
\phi_{LM}({\bf r})=\sum_k\hbox{$f_k\kern -1pt
{\displaystyle \int}\kern -0.3em d$}
\hat{\bf s}_kY_{LM}(\hat{\bf s}_k)
\varphi_{\mbox{\scriptsize{\bf s}}_k}^b({\bf r}),
\label{GCexpansion}
\end{equation}
where the gaussians are 0s h.o. functions of parameter $b$,
\beq
\varphi_{\mbox{\scriptsize{\bf s}}}^b({\bf r})=
\left({b \over \pi} \right)^{3/4}
\exp
\left\{-{b \over 2} ({\bf r}-{\bf s})^2
\right\}.
\eq
Accordingly, $\Psi$ will take the form $\Psi=\sum_kf_k\Psi_k$, with
$\Psi_k$ to be discussed below.

Let us assume that the two intrinsic states are single configurations
of h.o. states $\varphi^{\alpha}_{000}$ and $\varphi^{\beta}_{n_il_im_i}$,
($n_1,l_1,m_1,... ,n_A,l_A,m_A$), of parameters $\alpha$ and $\beta$,
respectively. If, in addition, the core is taken infinitely heavy,
$\Phi_{\rm A}$ reduces to a Slater determinant of
$\{\varphi^{\beta}_{n_il_im_i}\}$ and, by choosing $b=4\alpha$, so does
$\varphi_{{\mbox{\scriptsize{\bf s}}}_k}^b({\bf r})\Phi_a(\xi_a)$. The
corresponding $a+A$-particle state thus reads
\beqa
\Psi_k&=&
{\cal A}\{\varphi^{4\alpha}({\bf r}-{\bf s}_k)
\Phi_{\alpha}(\xi_{\alpha})\Phi_{\rm c}(\xi_{\rm c})\}
\label{eq:cmfactorizedclstate}\\
&=&[(a+A)!]^{-1/2}{\rm det}\{
\varphi^{\alpha}_{{\mbox{\scriptsize{\bf s}}}_k}({\bf r}_1)
\chi_1(1)...\varphi^{\alpha}_{{\mbox{\scriptsize{\bf s}}}_k}({\bf r}_a)
\chi_a(a)
\nonumber\\
&&\times\varphi^{\beta}_{n_1l_1m_1}({\bf r}_{a+1})\chi_{a+1}(a+1)...
\varphi^{\beta}_{n_Al_Am_A}({\bf r}_{a+A})\chi_{a+A}(a+A)\}.
\label{eq:clSlater}
\eqa
The coordinates ${\bf r}_i$ are those with respect to the total
c.m., but, the mass being infinity, the c.m. can be fixed at the
origin, so ${\bf r}_i$ may be interpreted as the usual s.p.
coordinates. The symbol $\chi_i(i)$ stands for a spin-isospin state.
The function $\Psi_k$ may be cast into an even more useful
form,
\beqa
\Psi_k&=&[(a+A)!]^{-1/2}{\rm det}\{
\psi_{{\mbox{\scriptsize{\bf s}}}_k}({\bf r}_1)\chi_1(1)...
\psi_{{\mbox{\scriptsize{\bf s}}}_k}({\bf r}_a)\chi_a(a)
\nonumber\\
&&\times\varphi^{\beta}_{n_1l_1m_1}({\bf r}_{a+1})\chi_{a+1}(a+1)...
\varphi^{\beta}_{n_Al_Am_A}({\bf r}_{a+A})\chi_{a+A}(a+A)\},
\label{eq:clortSlater}
\eqa
where
\beq
\psi_{\mbox{\scriptsize{\bf s}}}({\bf r}_j)=(1-{\cal P}_j)
\varphi_{\mbox{\scriptsize{\bf s}}}^{\alpha}({\bf r}_j)
=\varphi_{\mbox{\scriptsize{\bf s}}}^{\alpha}({\bf r}_j)-
\sum_{i=1}^{A}  \varphi_{n_il_im_i}^{\beta} ({\bf r}_j)
\langle \varphi_{n_il_im_i}^{\beta} \vert
\varphi_{\mbox{\scriptsize{\bf s}}}^{\alpha} \rangle,
\label{eq:projection}
\eq
with ${\cal P}_j=\sum_{i}^A\varphi_{n_il_im_i}^{\beta}({\bf r}_j)
\langle \varphi_{n_il_im_i}^{\beta}\vert$. We call the s.p. states
$\psi_{\mbox{\scriptsize{\bf s}}}({\bf r})$ Pauli-projected s.p. orbits.
The function $\Psi_k$ of eq.~(\ref{eq:clortSlater}) is equal to that
of eq. (\ref{eq:clSlater}) because (\ref{eq:clortSlater})
is obtained from (\ref{eq:clSlater}) by adding linear
combinations of columns $a+1,...,a+A$ to columns $1,...,a$.

The advantage of using (\ref{eq:clortSlater}) rather than (\ref{eq:clSlater})
is that the s.p. states involved in eq.~(\ref{eq:clortSlater}) are all
orthogonal to each other. The matrix elements between two states of the type of
eq. (\ref{eq:clortSlater}) can be expressed in terms of the s.p. states more
easily, which helps to separate the problem of the valence nucleons from that
of the core~\cite{Varga2}. In particular, the two-body matrix elements of
Slater determinants built up of orthogonal s.p. states
can be expressed simply by the two-body matrix elements of the s.p.
states.

\vskip 24pt
\begin{center}
\S {\bf 3. Matrix elements}
\end{center}
\vskip 12pt

We shall consider the two-body matrix elements of the
form factor, $e^{-{1 \over 2}\vartheta({\bf r}_2-{\bf r}_1)^2}$,
of a gaussian two-body potential between the Pauli-projected s.p.
functions $\psi_{\mbox{\scriptsize{\bf s}}}({\bf r})$:
\beqa
&&\langle
\psi_{\mbox{\scriptsize{\bf s}}} ({\bf r}_1)
\psi_{\mbox{\scriptsize{\bf s}}} ({\bf r}_2)
\left\vert
e^{-{1 \over 2} \vartheta \left( {\bf r}_2 - {\bf r}_1 \right)^2}
\right\vert
\psi_{{\mbox{\scriptsize{\bf s}}}'} ({\bf r}_1)
\psi_{{\mbox{\scriptsize{\bf s}}}'} ({\bf r}_2)
\rangle \nonumber \\
&=&\langle
\left(1-{\cal P}_1 \right)
\varphi_{\mbox{\scriptsize{\bf s}}}^{\alpha} ({\bf r}_1)
\left(1-{\cal P}_2 \right)
\varphi_{\mbox{\scriptsize{\bf s}}}^{\alpha} ({\bf r}_2)
\left\vert
e^{-{1 \over 2} \vartheta \left( {\bf r}_2 - {\bf r}_1 \right)^2}
\right\vert
\left(1 - {\cal P}_1 \right)
\varphi_{{\mbox{\scriptsize{\bf s}}}'}^{\alpha} ({\bf r}_1)
\left(1 - {\cal P}_2 \right)
\varphi_{{\mbox{\scriptsize{\bf s}}}'}^{\alpha} ({\bf r}_2)
\label{me}
\rangle .
\eqa
It should be noted that the s.p. states in the bra and ket have no
definite angular momenta, but a state of good angular momentum can be
easily constructed from any of $\psi_{\mbox{\scriptsize{\bf s}}}$
by taking ${\int\kern -0.2em d}\hat{\bf s}Y_{lm}(\hat{\bf s})
\psi_{\mbox{\scriptsize{\bf s}}}$. Since,
however, the projection of the total angular momentum in
eq.~(\ref{GCexpansion}) also involves the s.p. states other than
those in the two-particle matrix element (\ref{me}), it is more
natural to include this projection at a later stage, which is out of
the scope of this paper. Moreover, as is obvious from
eq.~(\ref{GCexpansion}), the projection of the full orbital angular
momentum only requires a single operation by
${\int\kern -0.2em d}\hat{\bf s}Y_{\cal LM}(\hat{\bf s})$.
The only thing we shall do in this paper to
facilitate angular-momentum projecting is that we shall express
the matrix element (\ref{me}) in terms of $s$, $\hat{\bf s}$, $s'$ and
$\hat{\bf s}'$.

By letting $1-{\cal P}_i$ act term by term, this matrix element is written as
a sum of 16 terms. Each of these terms contains two-body matrix elements of
different combinations of h.o. and gaussian wave functions.
To write the two-body matrix elements of the orbits
$\varphi^{\alpha}_{\mbox{\scriptsize{\bf s}}}$
and ${\cal P}\varphi^{\alpha}_{\mbox{\scriptsize{\bf s}}}$
as special cases of a general formula, we introduce the notation
\beq
M(q_{1}q_{2} \vert q_{1}'q_{2}') =
\langle {\cal P}_{1}^{q_1}
        \varphi_{\mbox{\scriptsize{\bf s}}}^{\alpha}({\bf r}_1)
        {\cal P}_{2}^{q_2}
	\varphi_{\mbox{\scriptsize{\bf s}}}^{\alpha}({\bf r}_2)
\left\vert
e^{-{1 \over 2} \vartheta \left( {\bf r}_2 - {\bf r}_1 \right)^2}
\right\vert
        {\cal P}_{1}^{q_1}
	\varphi_{\mbox{\scriptsize{\bf s}}'}^{\alpha}({\bf r}_1)
        {\cal P}_{2}^{q_2}
	\varphi_{\mbox{\scriptsize{\bf s}}'}^{\alpha}({\bf r}_2)
\rangle,
\label{eq:1}
\eq
which runs over all terms when each quotient $q_i$ runs over 0 and 1.
With these, (\ref{me}) can be expressed as
\beqa
& &
\langle
\psi_{\mbox{\scriptsize{\bf s}}} ({\bf r}_1)
\psi_{\mbox{\scriptsize{\bf s}}} ({\bf r}_2)
\left\vert
e^{-{1 \over 2} \vartheta \left( {\bf r}_2 - {\bf r}_1 \right)^2}
\right\vert
\psi_{{\mbox{\scriptsize{\bf s}}}'} ({\bf r}_1)
\psi_{{\mbox{\scriptsize{\bf s}}}'} ({\bf r}_2)
\rangle =M(00 \vert 00)-M(10 \vert 00) -M(01 \vert 00)
\nonumber \\
& &
-M(00 \vert 10)-M(00 \vert 01)
+M(11 \vert 00)+M(00 \vert 11)+M(10 \vert 10)+M(01 \vert 01)+M(10 \vert 01)
\nonumber \\
& &
+M(01 \vert 10)
-M(01 \vert 11)-M(10 \vert 11)-M(11 \vert 01)-M(11 \vert 10)
+M(11 \vert 11) .
\eqa
It is obvious that any two-body matrix elements that involve
shell-model orbits $\varphi_{n_il_im_i}^{\beta}$ instead of some of
the Pauli-projected excentric states
$\psi_{\mbox{\scriptsize{\bf s}}}$ are built up from the same
building blocks, but are simpler, and can be derived analogously.

The calculations are simplified by the following symmetry relations:
\beq
M(10 \vert 11)=M(01 \vert 11)  , \ \ \ \
M(11 \vert 10)=M(11 \vert 01)=M(01 \vert 11)' ,
\eq
\beq
M(00 \vert 10)=M(00 \vert 01)  , \ \ \ \
M(10 \vert 00)=M(01 \vert 00)=M(00 \vert 01)' ,
\eq
\beq
M(10 \vert 10)=M(01 \vert 01)' , \ \ \ \
M(11 \vert 00)=M(00 \vert 11)' , \ \ \ \
M(10 \vert 01)=M(01 \vert 10)' ,
\eq
where $M(q_1 q_2 \vert q_1' q_2')'$ differs from
$M(q_1 q_2 \vert q_1' q_2')'$ in that the parameters ${\bf s}$
and ${\bf s'}$ are exchanged.

It is convenient to factorize the h.o. function as
\beq
\varphi_{nlm}^{\beta}({\bf r})=e^{-{\beta \over 2} r^2}
{\hat \varphi}_{nlm}^{\beta}({\bf r}).
\eq
With the introduction of $A_{nl}(s)$ through
\beq
\langle \varphi_{\mbox{\scriptsize{\bf s}}}^{\alpha}
\vert\varphi_{nlm}^{\beta}\rangle=A_{nl}(s)Y_{lm}(\hat{\bf s}),
\label{eq:radp}
\eq
the matrix elements of eq. (\ref{eq:1}) take the form
\beqa
&&M(q_{1}q_{2} \vert q_{1}'q_{2}')=
{\cal N}_{q_1}(s){\cal N}_{q_2}(s){\cal N}_{q_1'}(s'){\cal N}_{q_2'}(s')
\nonumber\\
&&\times
\sum_{\lbrace n_{i}l_{i}m_{i} \rbrace_{q_i} \atop
       \lbrace n_{i}'l_{i}'m_{i}' \rbrace_{q'_i}}
A_{n_1 l_1} (s) Y_{l_1 m_1}({\hat {\bf s}})
A_{n_2 l_2} (s) Y_{l_2 m_2}({\hat {\bf s}})
A_{n_1' l_1'} (s') Y_{l_1' m_1'}({\hat {\bf s}}')^{\ast}
A_{n_2' l_2'} (s') Y_{l_2' m_2'}({\hat {\bf s}}')^{\ast} \nonumber \\
&&\times
\int d{\bf r}_1\int d{\bf r}_2
{\hat \varphi}_{n_1 l_1 m_1}^{\beta} ({\bf r}_1)^{\ast}
{\hat \varphi}_{n_2 l_2 m_2}^{\beta} ({\bf r}_2)^{\ast}
e^{-{1 \over 2}
\sum_{i=1}^{2} \left(
\alpha_i \left( {\bf r}_i - {\mbox{\scriptsize{\bf s}}} \right)^2
+ \beta_i  r_{i}^{2} \right)}
\nonumber\\
&&\times
e^{-{1 \over 2} \vartheta \left( {\bf r}_2 - {\bf r}_1 \right)^2}
{\hat \varphi}_{n_1' l_1' m_1'}^{\beta} ({\bf r}_1)
{\hat \varphi}_{n_2' l_2' m_2'}^{\beta} ({\bf r}_2)
e^{-{1 \over 2}
\sum_{i=1}^{2} \left(
\alpha_i' \left( {\bf r}_i - {\mbox{\scriptsize{\bf s}}'} \right)^2
+ \beta_i'  r_{i}^{2} \right)},
\eqa
with the notations $\alpha_i=(1-q_i)\alpha$, $\beta_i=q_i\beta$
and
\beq
{\cal N}_{q}(s)=\left[
\left( {\alpha \over \beta} \right)^{3/4}
{1 \over \langle \varphi_{\mbox{\scriptsize{\bf s}}}^{\alpha}
\vert \varphi_{000}^{\beta} \rangle}
\right]^{1-q}.
\eq
As to the summation limits,
if $q_i=1$ then $\{n_il_im_i\}$ runs over the
filled orbits, but, if $q_i=0$, then there is just one term,
$\{n_i=0, l_i=0, m_i=0\}$. For $A_{nl}$, an explicit formula is given
in eq. (\ref{eq:anl}) of the Appendix.

It is useful to couple the angular momenta of the h.o.
wave functions of the same arguments and make use of the formula
\cite{talman}
\beq
[{\hat \varphi}_{n_1 l_1}^{\beta}({\bf r}) {\hat \varphi}_{n_2 l_2}^{\beta}
({\bf r})]_{lm}=\left({\beta \over \pi }\right)^{3/4}
\sum_{n} T_{n_1 l_1 n_2 l_2}^{nl} {\hat \varphi}_{nlm}^{\beta}({\bf r})
\eq
(the coefficient $T_{n_1 l_1 n_2 l_2}^{nl}$ is given in eq.~(\ref{Tnlnl})
of the Appendix), to obtain
\[
M(q_{1}q_{2} \vert  q_{1}'q_{2}')=
\left({\beta \over \pi} \right)^{3/2}
{\cal N}_{q_1}(s){\cal N}_{q_2}(s){\cal N}_{q_1'}(s'){\cal N}_{q_2'}(s')
\sum_{\lbrace n_{i}l_{i} \rbrace_{q_i} \atop
       \lbrace n_{i}'l_{i}' \rbrace_{q'_i}}
A_{n_1 l_1}(s) A_{n_2 l_2}(s) A_{n_1' l_1'}(s') A_{n_2'l_2'}(s')
\]
\beq
\times
\sum_{\lambda \lambda' L M}
\left[ \left[ Y_{l_1 }({\hat {\bf s}}) Y_{l_1' }({\hat {\bf s}}')
       \right]_{\lambda}
       \left[ Y_{l_2 } ({\hat {\bf s}})  Y_{l_2' }({\hat {\bf s}}')
       \right]_{\lambda'} \right]_{LM}^{\ast}
\sum_{N N'}
T_{n_1 l_1 n_1' l_1'}^{N \lambda}
T_{n_2 l_2 n_2' l_2'}^{N' \lambda'}
I^{N N'}_{(\lambda \lambda')LM},
\label{eq:2}
\eq
with
\beqa
I^{N N'}_{(\lambda \lambda')LM}&=&
\int d{\bf r}_1 \int d{\bf r}_2
\left[
{\hat \varphi}_{N \lambda}^{\beta}
{\hat \varphi}_{N'\lambda'}^{\beta}
\right]_{LM}
\nonumber\\
&\times&
e^{-{1 \over 2}\sum_{i=1}^{2} \left(
\alpha_i \left( {\bf r}_i - {\bf s} \right)^2
+ \beta_i  r_{i}^{2} \right)}
e^{-{1 \over 2} \vartheta \left( {\bf r}_2 - {\bf r}_1 \right)^2}
e^{-{1 \over 2}
\sum_{i=1}^{2} \left(
\alpha_i' \left( {\bf r}_i - {\bf s'} \right)^2
+ \beta_i'  r_{i}^{2} \right)}.
\label{defofI}
\eqa
The summations over the angular momenta are restricted by the
well-known rules, i.e. $\vert l_1-l_1' \vert \le \lambda \le l_1+l_1'$,
and $M=-L,...,L$. The coefficient
$T^{nl}_{n_1 l_1 n_2 l_2}$ is non-zero if a triangle
inequality (see (\ref{eq:trian})
in the Appendix) of the quantum numbers is satisfied.
In eq. (\ref{eq:2}) the spherical harmonics can be recoupled so that
those belonging to the same argument can be combined:
\beqa
\left[ \left[ Y_{l_1 }({\hat {\bf s}}) Y_{l_1' }({\hat {\bf s}}')
       \right]_{\lambda}
       \left[ Y_{l_2 } ({\hat {\bf s}})  Y_{l_2' }({\hat {\bf s}}')
       \right]_{\lambda'} \right]_{LM}^{\ast}=
{\hat \lambda} {\hat \lambda'} \sum_{l l'}  {\hat l} {\hat l'}
\left(
\begin{array}{ccc}
l_1 & l_1' & \lambda \\
l_2 & l_2' & \lambda' \\
 l        &         l' & L
\end{array}
\right)
S_{l_1 l_2}^{l} S_{l_1' l_2'}^{l'}
\left[ Y_{l }({\hat {\bf s}}) Y_{l' }({\hat {\bf s}}')
       \right]_{LM}^{\ast}
\label{eq:3}
\eqa
where ${\hat J}=\sqrt{2J+1}$ and
\beq
S_{l_1 l_2}^{l}={{\hat l_1} {\hat l_2} \over \sqrt{4\pi} {\hat l}}
\langle l_1 0 l_2 0 \vert l 0 \rangle . \nonumber
\eq
In \S 4 we shall see that at least one of the angular
momenta in the 9j-symbol is zero in all practical cases,
therefore (\ref{eq:3}) reduces to a
simpler form.

The integral $I^{N N'}_{(\lambda \lambda')LM}$
can be determined using the the generating function of
${\hat \varphi}_{nlm}^{\beta}$ \cite{talman},
\beq
g({\bf p})=e^{2\sqrt{\beta}{\bf p}{\bf r}-p^2} .
\eq
With this, ${\hat \varphi}_{nlm}^{\beta}$  can be expressed as
\beqa
{\hat \varphi}_{nlm}^{\beta}=
\left( {\beta \over \pi} \right)^{3/4}
{c_{nl} \over (2n+l)!}
{d^{2n+l}\over dp^{2n+l} }
\left.\left( \int d{\hat {\bf p}}
Y_{lm}({\hat{\bf p}}) g({\bf p}) \right)
\right\vert_{p=0},
\nonumber\\
c_{nl}=(-1)^n \left({ n! (2n+2l+1)!!
\over 4 \pi 2^{n+l}}\right)^{1/2} .
\eqa
Thus the integral  $I^{N N'}_{(\lambda \lambda')LM}$ can be determined
from the matrix element of the generator function
\beq
G({\bf p},{\bf p'})=
\int d{\bf r}_1 \int d{\bf r}_2
g({\bf p})
e^{-{1 \over 2}
\sum_{i=1}^{2} \left(
\alpha_i \left( {\bf r}_i - {\bf s} \right)^2
+ \beta_i  r_{i}^{2} \right)}
e^{-{1 \over 2} \vartheta \left( {\bf r}_2 - {\bf r}_1
\right)^2}
e^{-{1 \over 2}
\sum_{i=1}^{2} \left(
\alpha_i' \left( {\bf r}_i - {\bf s'} \right)^2
+ \beta_i'  r_{i}^{2} \right)}
g({\bf p}') ,
\eq
as
\beqa
I^{N N'}_{(\lambda \lambda')LM}&=&\left({\beta \over \pi}\right)^{3/2}
{c_{N\lambda} c_{N' \lambda'} \over
(2N+\lambda)! (2N'+\lambda')!}
\nonumber\\
&\times&{\partial^{2N+\lambda+2N'+\lambda'} \over
\partial p^{2N+\lambda} \partial {p'}^{2N'+\lambda'}}
\left.\left(
\int d{\hat {\bf p}}
\int d{\hat {\bf p}}'
\left[Y_{\lambda \mu}({\hat {\bf p}})
      Y_{\lambda' \mu'}({\hat {\bf p}}') \right]_{LM}
      G({\bf p},{\bf p'}) \right) \right\vert_{p=0 \atop p'=0}\ .
\label{eq:4}
\eqa
With the abbreviations
\beq
\begin{array}{ccc}
a_1={1 \over 2}(\alpha_1+\alpha_1'+\beta_1+\beta_1'+\vartheta) ,&
a_2={1 \over 2}(\alpha_2+\alpha_2'+\beta_2+\beta_2'+\vartheta) ,
\\
d=4 a_{1} a_{2} - \vartheta^2 , \ \ \ \ \
A_1={a_2 \over d} , \ \ \ \ \
A_2={a_1 \over d} ,
&
{\bf P}=2\sqrt{\beta}{\bf p} , \ \ \ \ \
{\bf P'}=2\sqrt{\beta}{\bf p'} , \\
B_1={1 \over 4 \beta}-A_1 , \ \ \ \ \
B_2={1 \over 4 \beta}-A_2 , \ \ \ \ \
B={\theta \over d} ,
&
v=2 A_1\alpha_1\alpha_1'+2A_2\alpha_2\alpha_2'+
B(\alpha_1\alpha_2'+\alpha_1'\alpha_2) ,
\\
u={1\over2}
(\alpha_1+\alpha_2)-A_1\alpha_1^2-A_2\alpha_2^2-B\alpha_1\alpha_2
 , &
u'={1\over2}
(\alpha_1'+\alpha_2')-A_1{\alpha_1'}^2-A_2{\alpha_2'}^2-
B\alpha_1'\alpha_2' ,
\\
t=(2A_1\alpha_1 +B\alpha_2) , \ \ \ \ \
r=(2A_1\alpha_1' +B\alpha_2') , &
t'=(2A_2\alpha_2 +B\alpha_1)  , \ \ \ \ \
r'=(2A_2\alpha_2' +B\alpha_1') , \\
{\bf Q}=t {\bf s} +  r {\bf s'}  , &
{\bf Q'}=t' {\bf s} + r' {\bf s'} ,
\end{array}
\eq
the double integral $G({\bf p},{\bf p'})$ is expressible as
\beq
G({\bf p},{\bf p'})=
\left( {4 \pi^2 \over d} \right)^{3/2}
\exp\lbrace -u s^2 - u' s'^2 + v {\bf ss'}
-B_1 P^2 - B_2 P'^2
+ B {\bf PP'} + {\bf PQ} + {\bf P'Q'}
\rbrace .
\eq

The explicit forms of these coefficients in the exponent
for different values of $q_i$
and $q_i'$ are collected in table 1. To be able to carry out
the operations prescribed by (\ref{eq:4}), we
expand the exponentials containing ${\bf P}$ and ${\bf P'}$
into the power series
\beqa
G({\bf p},{\bf p}')&=&
\left( {4 \pi^2 \over d} \right)^{3/2}
\exp\lbrace -u{\bf s}^2 -u'{\bf s'}^2+v{\bf ss'}
\rbrace
\sum_{\nu_1 \nu_2 \mu_1 \mu_2 \kappa}
{(-1)^{\nu_1+\nu_2} \over \nu_{1}! \nu_{2}! \mu_{1}! \mu_{2}! \kappa !}
p^{2\nu_1+\mu_1+\kappa}
p'^{2\nu_2+\mu_2+\kappa}
\nonumber \\
&&\times
B_1^{\nu_1}
B_2^{\nu_2}
\left(2\sqrt{\beta}\right)^{2(\nu_1+\nu_2+\kappa)+\mu_1+\mu_2}
B^{\kappa}
Q^{\mu_1}
{Q'}^{\mu_2}
\sum_{\lambda_1 \lambda_2 l}
C_{\mu_{1}\lambda_1}
C_{\mu_{2}\lambda_2}
C_{\kappa l} \nonumber \\
&&\times
\sum_{m_1 m_2 m}
Y_{\lambda_1 m_1}({\hat {\bf p}})^{\ast} Y_{\lambda_1 m_1}({\hat {\bf Q}})
Y_{\lambda_2 m_2}({\hat {\bf p}}')^{\ast} Y_{\lambda_2 m_2}({\hat {\bf Q}}')
Y_{l m}({\hat {\bf p}}) Y_{l m}({\hat {\bf p}}')^{\ast} ,
\label{eq:5}
\eqa
where we used the expression
\beq
({\bf a b})^n=a^n b^n
{\sum_{l(n)}}
C_{nl}
\sum_{m=-l}^{l}
Y_{l m}({\hat {\bf a}})Y_{l m}({\hat {\bf b}})^{\ast}, \ \ \ \ \ \
C_{nl}={4 \pi n! \over (n-l)!! (n+l+1)!!}  .
\label{eq:cnl}
\eq
In this formula ${\sum_{l(n)}}$ stands for a summation over
$l=0,2, ..., n-2,n$ if $n$ is even, and $l=1,3, ..., n-2,n$
if $n$ is odd.

It is straightforward to recouple the spherical harmonics in
(\ref{eq:5}):
\beqa
&&\sum_{m_1 m_2 m}
Y_{\lambda_1 m_1}({\hat {\bf p}})^{\ast} Y_{\lambda_1 m_1}({\hat {\bf Q}})
Y_{\lambda_2 m_2}({\hat {\bf p}}')^{\ast} Y_{\lambda_2 m_2}({\hat {\bf Q}}')
Y_{l m}({\hat {\bf p}}) Y_{l m}({\hat {\bf p}}')^{\ast}
\nonumber\\
&=&\sum_{\lambda \lambda' L M} (-1)^{l} {\hat \lambda} {\hat \lambda'}
W(\lambda_1 L l \lambda'; \lambda_2 \lambda)
S_{\lambda_1 l}^{\lambda} S_{\lambda_2 l}^{\lambda'}
\left[
Y_{\lambda}({\hat {\bf p}}) Y_{\lambda'}({\hat {\bf p}}')
\right]_{LM}^{\ast}
\left[
Y_{\lambda_1}({\hat {\bf Q}}) Y_{\lambda_2}({\hat {\bf Q}}')
\right]_{LM} .
\label{eq:6}
\eqa
After substituting (\ref{eq:6}) into (\ref{eq:5}) and inserting
(\ref{eq:5}) in (\ref{eq:4}) the derivations and
the integrations
in (\ref{eq:4}) can be carried out, with the result
\beq
I^{N N'}_{(\lambda \lambda')LM} =
\exp\lbrace -u{\bf s}^2-u'{\bf s'}^2+v{\bf ss'}\rbrace
\sum_{\mu_1=0}^{2N+\lambda}
\sum_{\mu_2=0}^{2N'+\lambda'}
\sum_{\lambda_1(\mu_1)}
\sum_{\lambda_2(\mu_2)}
G^{(N\lambda N'\lambda')L}_{\mu_1 \lambda_1 \mu_2 \lambda_2}
Q^{\mu_1}
{Q'}^{\mu_2}
\left[
Y_{\lambda_1}({\hat {\bf Q}}) Y_{\lambda_2}({\hat {\bf Q}}')
\right]_{LM} ,
\label{eq:inl}
\eq
where
\beqa
G^{(N\lambda N'\lambda')L}_{\mu_1 \lambda_1 \mu_2 \lambda_2}&=&
\left( {4\pi\beta\over d}\right)^{3/2} {\hat\lambda}{\hat \lambda'}
\left(2\sqrt{\beta}\right)^{2N+\lambda+2N'+\lambda'}
c_{N \lambda} c_{N' \lambda'}
C_{\mu_1 \lambda_1} C_{\mu_2 \lambda_2}
\nonumber\\
&\times&\sum_{\nu_1 \nu_2 \kappa}
{(-1)^{\nu_1+\nu_2}
\over \nu_{1}! \nu_{2}! \mu_{1}! \mu_{2}! \kappa !}
B_1^{\nu_1}
B_2^{\nu_2}
B^{\kappa}
\sum_{l}(-1)^{l} W(\lambda_1 L l \lambda'; \lambda_2 \lambda)
C_{\kappa l}S_{\lambda_1 l}^{\lambda} S_{\lambda_2 l}^{\lambda'} ,
\label{eq:7}
\eqa
and the summation indices should fulfil the constraints
\beq
2\nu_1 +\mu_1+\kappa=2N+\lambda ,\ \ \ \ \ \
2\nu_2 +\mu_2+\kappa=2N'+\lambda' .
\label{eq:cons}
\eq
The vectors ${\bf Q}$ and ${\bf Q'}$ are linear
combinations of ${\bf s}$ and ${\bf s'}$. To facilitate the
angular-momentum projection,
it is useful to expand the ${\bf Q}$- and ${\bf Q}'$-dependent
factor of eq.~(\ref{eq:inl}) in terms of
${\bf s}$ and ${\bf s'}$:
\beqa
&&Q^{\mu_1}Q'^{\mu_2}\left[Y_{\lambda_1}({\hat {\bf Q}}) Y_{\lambda_2}({\hat
{\bf Q}}') \right]_{LM}
\nonumber\\
&=& \sum_{k_1=0}^{\mu_1}
    \sum_{k_2=0}^{\mu_2}
t^{k_1}r^{\mu_1-k_1}t'^{k_2}r'^{\mu_2-k_2}
s^{k_1+k_2}s'^{\mu_1+\mu_2-k_1-k_2}
\sum_{\lambda_1' \lambda_2'}
F^{\mu_1 k_1 \mu_2 k_2}_{\lambda_1 \lambda_2 \lambda_1' \lambda_2' L}
\left[Y_{\lambda_1'}({\hat {\bf s}}) Y_{\lambda_2'}({\hat
{\bf s}}') \right]_{LM} ,
\label{eq:ff}
\eqa
where the coefficients
$F^{\mu_1 k_1 \mu_2 k_2}_{\lambda_1 \lambda_2 \lambda_1' \lambda_2' L}$
are given in eq.~(\ref{eq:f}) of the Appendix.
One can verify this expansion immediately by using the well-known
expression \cite{rose}
\beq
(a+b)^n Y_{lm}({\widehat {\bf a+b}})=\sum_{k=0}^{n}
a^k b^{n-k} \sum_{\lambda_1(k)}\sum_{\lambda_2(n-k)}
H^{n k}_{\lambda_1 \lambda_2 l}
\left[ Y_{\lambda_1}({\hat {\bf a}}) Y_{\lambda_2}({\hat {\bf b}})\right]_{lm}
\label{eq:hh}
\eq
where $H^{n k}_{\lambda_1 \lambda_2 l}$ is given in eq.~(\ref{eq:h})
of the Appendix.

With  (\ref{eq:inl}) and (\ref{eq:ff})
substituted into eq.~(\ref{eq:2}), our final expression for
$M(q_1 q_2 \vert q_1' q_2')$ becomes
\beqa
&&M(q_{1} q_{2} \vert  q_{1}' q_{2}')=\left({\beta \over \pi} \right)^{3/2}
{\cal N}_{q_1}(s){\cal N}_{q_2}(s){\cal N}_{q_1'}(s'){\cal N}_{q_2'}(s')
\exp\lbrace -u{\bf s}^2-u'{\bf s'}^2+v{\bf ss'}\rbrace
\nonumber\\
&&\times
\sum_{\lbrace n_{i}l_{i} \rbrace_{q_i} \atop
       \lbrace n_{i}'l_{i}' \rbrace_{q'_i}}
A_{n_1 l_1}(s) A_{n_2 l_2}(s) A_{n_1' l_1'}(s') A_{n_2'l_2'}(s')
\sum_{\lambda \lambda'}
\sum_{l l'}
\sum_{L}(-1)^L{\hat L}^2
{\hat l}{\hat l}'{\hat \lambda} {\hat \lambda'}
\left(
\begin{array}{ccc}
\l_1 & \l_1' & \lambda \\
\l_2 & \l_2' & \lambda' \\
 l        &         l' & L
\end{array}
\right)
S_{l_1 l_2}^{l} S_{l_1' l_2'}^{l'}
\nonumber \\
& &\times\sum_{N N'}
T_{n_1 l_1 n_1' l_1'}^{N \lambda}
T_{n_2 l_2 n_2' l_2'}^{N' \lambda'}
\sum_{\mu_1=0}^{2N+\lambda}
\sum_{\mu_2=0}^{2N'+\lambda'}
\sum_{\lambda_1(\mu_1)}
\sum_{\lambda_2(\mu_2)}
G^{(N\lambda N'\lambda')L}_{\mu_1 \lambda_1 \mu_2 \lambda_2}
\nonumber \\
& &\times
\sum_{k_1}^{\mu_1}
\sum_{k_2}^{\mu_2}
t^{k_1}
r^{\mu_1-k_1}
t'^{k_2}
r'^{\mu_2-k_2}
s^{k_1+k_2}
s'^{\mu_1+\mu_2-k_1-k_2}
\nonumber\\
& &\times
\sum_{\lambda_1' \lambda_2'}
F^{\mu_1 k_1 \mu_2 k_2}_{\lambda_1 \lambda_2 \lambda_1' \lambda_2' L}
\sum_{L'} (-1)^{L'}
W(l l' \lambda_1' \lambda_2';L L')
S_{l \lambda_1'}^{L'} S_{l' \lambda_2'}^{L'} \sum_{M'}
Y_{L' M'}({\hat {\bf s}}) Y_{L' M'}({\hat {\bf s}}')^{\ast} .
\label{eq:main}
\eqa

\vskip 24pt
\begin{center}
\S {\bf 4. Special cases}
\end{center}
\vskip 12pt

In this section we substitute the values of $q_i$ and $q_i'$
into  eq. (\ref{eq:main}) and determine the concrete forms of  $M(q_1 q_2 \vert
q_1' q_2')$ .

(1) $M(11 \vert 11)$. This case reduces to combinations of
two-body matrix elements of h.o. functions.
These matrix elements are extensively used in shell-model
calculations \cite{Lawson}.
{}From table 1 we see that $t=t'=r=r'=0$, i.e.
${\bf Q} =0, {\bf Q}'=0$ , $u=u'=v=0$,
and then only the terms with
$\mu_i=k_i=\lambda_i=0 , (i=1,2)$ survive in
equation (\ref{eq:main}). Moreover, as
\beq
F^{0000}_{\lambda_1 \lambda_2 \lambda_1' \lambda_2' L}=
\delta_{\lambda_1 0}
\delta_{\lambda_2 0}
\delta_{\lambda_1' 0}
\delta_{\lambda_2' 0}
\delta_{L 0}
\eq
the the Racah, the 9j-symbol and the $S_{l l'}^L$ coefficients in equation
(\ref{eq:main}) are reduced as:
\beq
W(ll00;0L')=
{(-1)^{l-L'} \over {\hat l}} ,
\eq
\beq
\left(
\begin{array}{ccc}
\l_1 & \l_1' & \lambda \\
\l_2 & \l_2' & \lambda' \\
 l        &         l' & 0
\end{array}
\right)
=
\delta_{\lambda \lambda'}
\delta_{l l'}
{(-1)^{\lambda+l+l_1+l_2'}\over {\hat \lambda} {\hat l}}
W(l_1 l_1' l_2 l_2';\lambda l)
\eq
and
\beq
S_{l0}^{L'}={1 \over \sqrt{4\pi}} \delta_{lL'} ,
\ \ \ \ \ \ \ \ \ \
S_{l'0}^{L'}={1 \over \sqrt{4\pi}} \delta_{l'L'} .
\eq
By using these results, eq. (\ref{eq:main}) reads as
\beqa
M(1 1 \vert 1 1) = & &
 \left({\beta \over \pi} \right)^{3/2}
\sum_{\lbrace n_{i}l_{i} \rbrace \atop
       \lbrace n_{i}'l_{i}' \rbrace}
A_{n_1 l_1}(s) A_{n_2 l_2}(s) A_{n_1' l_1'}(s') A_{n_2'l_2'}(s')
\sum_{l \lambda} {\hat \lambda} (-1)^l {1 \over 4 \pi}
W(l_1 l_1' l_2 l_2'; \lambda l)
\nonumber \\
& &
\times S_{l_1 l_2}^{l} S_{l_1' l_2'}^{l'}
\sum_{m}
Y_{lm}({\hat {\bf s}}) Y_{lm}({\hat {\bf s}}')^{\ast}
 \sum_{NN'}
T_{n_1 l_1 n_1' l_1'}^{N \lambda}
T_{n_2 l_2 n_2' l_2'}^{N' \lambda}
G^{(N\lambda N'\lambda)0}_{0000} .
\nonumber
\eqa

In this case the coefficient $G^{(N\lambda N'\lambda)0}_{0000}$
can be expressed in a simpler form.
{}From eq. (\ref{eq:cons}) it follows that
$\kappa=2N+\lambda-2\nu_1$ and $\nu_2=N'-N+\nu_1$, and then
by substituting $B_1$, $B_2$ and $B$ (see table 1) into
eq. (\ref{eq:7}), reducing the Racah and the $S_{ll'}^{L}$
coefficients, $G^{(N\lambda N'\lambda)0}_{0 0 0 0}$ is given by
\beqa
G^{(N\lambda N'\lambda)0}_{0 0 0 0}= & &
(4\pi)^2 \left( {4 \pi \beta \over d} \right)^{3/2}
c_{N \lambda} c_{N' \lambda}
\left(2\sqrt{\beta}\right)^{2N+\lambda+2N'+\lambda}
(-1)^{N+N'+\lambda} {\hat \lambda}
\left( {\vartheta/2 \over \vartheta+\beta} \right)^{N+N'+\lambda}
\nonumber \\
& &
\sum_{\nu_1}
{2^{N+\lambda-\nu_1} \over
\nu_1 ! (N'-N+\nu_1)! (N-\nu_1)! (2N+2\lambda-2\nu_1+1)!! } .
\eqa
The summation over $\nu_1$ can be carried out \cite{talman} and our final
expression is
\beqa
G^{(N\lambda N'\lambda)0}_{0000} = & &
\left( {4 \beta \pi \over d} \right)^{3/2}
c_{N \lambda} c_{N' \lambda}
\left( {-\vartheta/2 \over \vartheta+\beta }\right)^{N+N'+\lambda}
(-1)^{\lambda} 2^{\lambda}  (4 \pi)^2 {\hat \lambda} \nonumber \\
& &
\times
{(2N+2N'+2\lambda+1)!! \over
N! N'! (2N+2\lambda+1)!! (2N'+2\lambda+1)!! } .
\eqa
(2) $M(01 \vert 11)$.
Now we do not have so simple expressions as in the previous case.
The first simplification is that
$r=r'=0$, i.e.
${\bf Q}=t{\bf s}$,
${\bf Q'}=t'{\bf s}$ and
$u'=v=0$
(see table 1).
The summation over $k_i (i=1,2)$ is then restricted to
$k_i=\mu_i (i=1,2)$. The coefficient
$F^{\mu_1 \mu_1 \mu_2 \mu_2}_{\lambda_1 \lambda_2 \lambda_1'
\lambda_2' L}$ can easily be evaluated:
\beq
F^{\mu_1 \mu_1 \mu_2 \mu_2}_{\lambda_1 \lambda_2 \lambda_1'
\lambda_2' L}=\sqrt{4\pi} \delta_{\lambda_1' L}
\delta_{\lambda_2' 0} S_{\lambda_1 \lambda_2}^L .
\eq
{}From this it follows that $\lambda_2'=0$, thus the Racah coefficient
and $S_{l' \lambda_2'}^{L'}$ in equation (\ref{eq:main}) can be
written as:
\beq
W(ll'L0;LL')={\delta_{l' L'} \over {\hat L } {\hat L'}} ,
\ \ \ \ \ \ \ \ \
S_{l' 0}^{L'}={1\over \sqrt{4 \pi}} \delta_{l' L'}
\eq

An other simplification to be used is that
when
$q_1=0$,  then $n_1=l_1=0$. From this it follows that the
coefficient $T_{n_1 l_1
n_1' l_1'}^{N \lambda}$  is
\beq
T_{0 0 n_1' l_1'}^{N \lambda}=\delta_{n_1' N} \delta_{l_1' \lambda}
\eq
and the 9j-symbol, similarly as in the first case,
can be expressed by a Racah coefficient.

Let us put all these together to find
\beqa
M(0 1 \vert 1 1) = & &
 \left({\alpha \beta \over \pi^2} \right)^{3/4}
\exp\lbrace -u s^2 \rbrace
\sum_{\lbrace n_{i}l_{i} \rbrace \atop
       \lbrace n_{i}'l_{i}' \rbrace}
A_{n_2 l_2}(s) A_{n_1' l_1'}(s') A_{n_2'l_2'}(s')
\sum_{L L' \lambda'} {\hat \lambda'} {\hat L} (-1)^{l_2}
W(L' l_2 L \lambda'; l_1' l_2)
\nonumber \\
& &
\times
S_{l_1' l_2'}^{L'}
S_{l_2 L}^{L'}
\sum_{m'}
Y_{L'm'}({\hat {\bf s}}) Y_{L'm'}({\hat {\bf s}}')^{\ast}
 \sum_{N'}
T_{n_2 l_2 n_2' l_2'}^{N' \lambda'}
\sum_{\mu_1 \mu_2 \lambda_1 \lambda_2}
G^{(n_1' l_1' N'\lambda')L}_{\mu_1 \lambda_1 \mu_2 \lambda_2}
t^{\mu_1} t'^{\mu_2} s^{\mu_1+\mu_2} S_{\lambda_1 \lambda_2}^{L} .
\nonumber
\eqa

(3) $M(0 1 \vert 1 0)$.
As we can see in Table 1,  all coefficients differ from zero.
There is one gaussian in the bra and one in the ket, that is
$q_1=q_2'=0$ and $n_1=l_1=n_2'=l_2'=0$. It follows that
\beq
T_{0 0 n_1' l_1'}^{N \lambda}=\delta_{n_1' N} \delta_{l_1'\lambda} ,
\ \ \ \ \ \ \ \ \ \
T_{n_2 l_2 0 0}^{N' \lambda'}=\delta_{n_2 N'} \delta_{l_2\lambda'} ,
\eq
and the 9j-symbol, as it contains two zero elements, can be substituted
by
\beq
\left(
\begin{array}{ccc}
0    & \l_1' & \lambda \\
\l_2 & 0      & \lambda' \\
 l        &         l' & L
\end{array}
\right)
=
{(-1)^{{l_1}'+l_2-L} \over {\hat l_2}^2 {\hat {l_1'}}^2}
\delta_{l_1' \lambda}
\delta_{l l_2}
\delta_{l' l_1'}
\delta_{\lambda' l_2}
\eq

That is all one can do in this case and thus equation (\ref{eq:main})
becomes
\beqa
M(0 1 \vert 1 0) = & &
\left({\alpha \over \pi} \right)^{3/2}
\exp\lbrace -u{\bf s}^2-u'{\bf s'}^2+v{\bf ss'}\rbrace
\sum_{n_{2}l_{2}  \atop  n_{1}'l_{1}' }
A_{n_2 l_2}(s) A_{n_1' l_1'}(s')
\sum_{L}
\sum_{\mu_1=0}^{2n_1'+l_1'}
\sum_{\mu_2=0}^{2n_2+l_2}
\nonumber \\
& &
\sum_{\lambda_1(\mu_1)}
\sum_{\lambda_2(\mu_2)}
G^{n_1' l_1' n_2 l_2}_{\mu_1 \lambda_1 \mu_2 \lambda_2}
\sum_{k_1=0}^{\mu_1}
\sum_{k_2=0}^{\mu_2}
t^{k_1}
r^{\mu_1-k_1}
t'^{k_2}
r'^{\mu_2-k_2}
s^{k_1+k_2}
s'^{\mu_1+\mu_2-k_1-k_2}
\sum_{\lambda_1' \lambda_2'}
F^{\mu_1 k_1 \mu_2 k_2}_{\lambda_1 \lambda_2 \lambda_1' \lambda_2' L}
\nonumber \\
& &
{\hat L}^2 \sum_{L'} (-1)^{l_1'+l_2+L'}
W(l_2 l_1' \lambda_1' \lambda_2';L L')
S_{l_2 \lambda_1'}^{L'} S_{l_1' \lambda_2'}^{L'} \sum_{M'}
Y_{L' M'}({\hat {\bf s}}) Y_{L' M'}({\hat {\bf s}}')^{\ast} .
\nonumber
\eqa

(4) $M(0 1 \vert 0 1)$.
Now  $q_1=q_1'=0$ and $n_1=l_1=n_1'=l_1'=0$, consequently
$T_{0000}^{N\lambda}=\delta_{N \lambda}$ and therefore $\mu_1=k_1=0$.
Then the coefficients will reduce to
\beq
F_{0 \lambda_2 \lambda_1' \lambda_2'L}^{0 0 \mu_2 k_2}=
{1 \over \sqrt{4\pi}} \delta_{L \lambda_2}
H_{\lambda_1' \lambda_2' \lambda_2}^{\mu_2 k_2} ,
\eq
and after repeating the steps used in the previous cases,
i.e. reducing the 9j-, Racah and S-symbols,
the expression (\ref{eq:main}) yields
\beqa
M(0 1 \vert 0 1) = & &
\left({\alpha^2 \beta \over \pi^3} \right)^{3/4}
\exp\lbrace -u{\bf s}^2-u'{\bf s'}^2+v{\bf ss'}\rbrace
\sum_{n_{2}l_{2} \atop n_{2}'l_{2}'}
A_{n_2 l_2}(s) A_{n_2' l_2'}(s')
\sum_{L}
\sum_{N'} T^{N' L}_{n_2 l_2 n_2'l_2'}
\sum_{\mu_2=0}^{2N'+L}
\nonumber \\
& &
\times
G^{(0 0 N'L)L}_{0 0  \mu_2 L}
\sum_{k_2}
t'^{k_2}
r'^{\mu_2-k_2}
s^{k_2}
s'^{\mu_2-k_2}
\sum_{\lambda_1' \lambda_2'}
H^{\mu_2 k_2}_{\lambda_1' \lambda_2' L}
{\hat L}^2 \sum_{L'} (-1)^{L+L'}
W(l_2 l_2' \lambda_1' \lambda_2';L L')
\nonumber \\
& &
\times
S_{l_2 \lambda_1'}^{L'} S_{l_2' \lambda_2'}^{L'} \sum_{M'}
Y_{L' M'}({\hat {\bf s}}) Y_{L' M'}({\hat {\bf s}}')^{\ast} ,
\nonumber
\eqa
where the coefficient G can be written as
\beq
G^{(0 0 N'\lambda')\lambda'}_{0 0 \mu_2 \lambda_2} =
\left( {4 \beta \pi \over d} \right)^{3/2}
c_{N' \lambda'}
\left(2\sqrt{\beta}\right)^{2N'+\lambda'}
\sum_{
2\nu_2 +\mu_2=2N'+\lambda'}
{(-1)^{\nu_2}
\over \nu_{2}! \mu_{2}! }
B_2^{\nu_2}
C_{\mu_2 \lambda'} ,
\eq

(5) $M(0 0 \vert 1 1)$. In this case both gaussians are in the bra, and
$u'=v=0$, $r=r'=0$, therefore
${\bf Q}=t{\bf s}, {\bf Q}=t'{\bf s}$ just as in case (2).
In addition, now $n_1=l_1=n_2=l_2=0$, so the result of case (2)
can be simplified further:
\beqa
M(0 0 \vert 1 1) = & &
\left({\alpha \over \pi} \right)^{3/2}
\exp\lbrace -u{\bf s}^2\rbrace
\sum_{ n_{1}'l_{1}' \atop n_{2}'l_{2}' }
A_{n_1' l_1'}(s) A_{n_2' l_2'}(s')
\sum_{L}
\sum_{\mu_1=0}^{2N+\lambda}
\sum_{\mu_2=0}^{2N'+\lambda'}
\sum_{\lambda_1}'^{\mu_1}
\sum_{\lambda_2}'^{\mu_2}
G^{(n_1' l_1' n_2' l_2')L}_{\mu_1 \lambda_1 \mu_2 \lambda_2}
\nonumber \\
& &
\times
t^{\mu_1}
t'^{\mu_2}
s^{\mu_1+\mu_2}
S_{\lambda_1 \lambda_2}^{L}
\sum_{M}
Y_{L M}({\hat {\bf s}}) Y_{L M}({\hat {\bf s}}')^{\ast} .
\nonumber
\eqa

(6) $M(0 0 \vert 0 1)$. Using the results of case (4), one finds
\beqa
M(0 0 \vert 0 1) = & &
\left({\alpha^3 \over \beta \pi^2 } \right)^{3/4}
\exp\lbrace -u{\bf s}^2-u'{\bf s'}^2+v{\bf ss'}\rbrace
\sum_{n_{2}'l_{2}'}
\sum_{\mu_2=0}^{2 n_2'+l_2'}
G^{(0 0 n_2' l_2')l_2'}_{0 0  \mu_2 l_2'}
\sum_{k_2}
t'^{k_2}
r'^{\mu_2-k_2}
s^{k_2}
s'^{\mu_2-k_2} \nonumber
\\
& &
\times
\sum_{\lambda_1' \lambda_2'}
H^{\mu_2 k_2}_{\lambda_1' \lambda_2' l_2'}
{{\hat l_2'} \over {\lambda_1'}} (-1)^{\lambda_2'}
S_{l_2' \lambda_1'}^{\lambda_2'}
\sum_{m_2}
Y_{\lambda_2' m_2}({\hat {\bf s}}) Y_{\lambda_2' m_2}
({\hat {\bf s}}')^{\ast} .
\nonumber
\eqa

(7) $M(0 0 \vert 0 0)$ This is the simplest case, overlap of
four gaussians is just
\beq
M(0 0 \vert 0 0)=
\left( {\alpha \over \alpha+\vartheta} \right)^{3/2}
\exp\lbrace -u{\bf s}^2-u'{\bf s'}^2+v{\bf ss'}\rbrace
\eq

\vskip 24pt
\begin{center}
\S {\bf 5. Conclusions}
\end{center}
\vskip 12pt

All summation in formula (\ref{eq:main}) are finite, so after angular
momentum projection it is straightforward to apply it, or its special
cases given in \S {\bf 4.}, in numerical calculations \cite{vk}. If, in any of
the states involved there are other nucleons in Pauli-projected
orbits, their overlaps may bring in factors of the form of ${\rm exp}
\lbrace w{\bf ss'}\rbrace$. Grouping these factors together with  ${\rm exp}
\lbrace v{\bf ss'}\rbrace$ of eq. \ref{eq:main}, the product has to be
expanded into its multipoles. The functions $Y_{\Lambda M}({\hat s})
Y_{\Lambda M}({\hat s'})^{\ast}$ appearing in this way should be coupled
with  $Y_{L'M'}({\hat s})Y_{L'M'}({\hat s'})^{\ast}$,
respectively. The infinite summation involved in the multipole
expansion is then eliminated by the angular momentum projection,
which can be achieved by the operations
$\int d{\hat s}Y_{\cal LM}^{\ast}({\hat s})$ and
$\int d{\hat s'}Y_{\cal L'M'}({\hat s'})$.

I am indebted to Prof. R.G. Lovas for carefull reading of the
manuscript.

This work was supported by the OTKA grant (No. 3010).

\vskip 24pt
\begin{center}
\S {\bf 6. Appendix}
\end{center}
\vskip 12pt

In this paper we used the standard definition of the
harmonic oscillator function (see e.g. \cite{talman})
\beq
{\varphi}_{nlm}^{\beta}({\bf r})=
\left( {\beta \over \pi} \right)^{3/4}
e^{-{\beta \over 2}r^2}
\left({ n! 4 \pi 2^{n+l} \over (2n+2l+1)!! } \right)^{1/2}
(\sqrt{\beta}r)^l L_{n}^{l+1/2}(\beta r^2)
Y_{lm}({\hat{\bf r}})
\eq
with $L_{n}^{l+1/2}(x)$ being associated the Laguerre polynomial
\beq
L_{n}^{l+1/2}(x)={(2n+2l+1)!! \over 2^{n}}
\sum_{s=0}^{n} {(-1)^s 2^{s}
\over (2s+2l+1)!! s! (n-s)! } x^{s} .
\eq

The radial part (see eq. (\ref{eq:radp}))  of the overlap
of a harmonic oscillator function
of width parameter $\beta$ and a gaussian of width parameter
$\alpha$ and of displacement vector $\bf s$ is
\beqa
A_{nl}=\sqrt{4\pi} & &
\left({n!(2n+2l+1)!! \over 2^{n+l}}\right)^{1/2}
\left({4\alpha \beta \over (\alpha+\beta)^2} \right)^{3/4}
e^{-{\alpha\beta \over 2(\alpha+\beta)^2}s^2}
\nonumber \\
& & \sum_{k=0}^{n} (-1)^k
{2^{k+l} \over k! (n-k)! (2k+2l+1)!! }
\left({\alpha - \beta \over \alpha+\beta} \right)^{n-k}
\left({\alpha \sqrt{\beta}s \over \alpha+\beta} \right)^{2k+l}
\label{eq:anl}
\eqa

The combination coefficients of product of two h. o.
functions of the  same argument ae given by \cite{talman}
\beqa
T_{nln'l'}^{NL} & = &
\left({ n! n'! 2^{N+L} (2n+2l+1)!! (2n'+2l'+1)!!
\over (2N+2L+1)!! 2^{n+l} 2^{n'+l'} N! } \right)^{1/2}
\langle l 0 l' 0 \vert L 0 \rangle
{{\hat l} {\hat l'} \over {\hat L}}
\sum_{s s'} (-1)^{s+s'}  \nonumber \\
& &
\times
{(2s+2s'+l+l'+L+1)!!(s+s'+{1 \over 2}(l+l'-L))!
\over
(2s+2l+1)!! (2s'+2l'+1)!! (n-s)! s! (n'-s')! s'!
(s+s'-N+{1 \over2}(l+l'-L))!} ,
\label{Tnlnl}
\eqa
where the quantum numbers must satify the triangular condition
\beq
\vert 2n+l-2n'-l \vert \le 2N+L \le 2n+l+2n'+l' .
\label{eq:trian}
\eq

The expansion coefficients in equation (\ref{eq:hh}) are
\beq
H^{n k}_{\lambda_1 \lambda_2 l}=\left(n \atop k \right)
{C_{k \lambda_1} C_{n-k \lambda_2} \over C_{n l}}
S_{\lambda_1 \lambda_2}^{l} .
\label{eq:h}
\eq

The detailed form of the coefficients of equation (\ref{eq:ff}) is
\beqa
F^{\mu_1 k_1 \mu_2 k_2}_{\lambda_1 \lambda_2 \lambda_1' \lambda_2' L}=
\sum_{l_1 l_2  l_1' l_2'}
H^{\mu_1 k_1}_{l_1 l_1' \lambda_1}
H^{\mu_2 k_2}_{l_2 l_2' \lambda_2}
S_{l_1 l_2}^{\lambda_1'}
S_{l_1' l_2'}^{\lambda_2'}
{\hat \lambda_1}
{\hat \lambda_2}
{\hat \lambda_1'}
{\hat \lambda_2'}
\left(
\begin{array}{ccc}
l_1 & l_1' & \lambda_1 \\
l_2 & l_2' & \lambda_2 \\
\lambda_1'& \lambda_2' & L
\end{array}
\right) .
\label{eq:f}
\eqa
The range of the summation variables are determined by those of
the coefficient $H^{n k}_{\lambda_1 \lambda_2 l}$ and those of the
$9j$ symbol.

{}

\begin{thebibliography}{99}
\bibitem{hori} H. Horiuchi, Prog. Theor. Phys. Suppl. {\bf 62}
               (1977) 90.

\bibitem{kami} M. Kamimura, Prog. Theor. Phys. Suppl. {\bf 62}
               (1977)

\bibitem{Wild} T. Steinmayer, W. S\"unkel, and K. Wildermuth,
Phys. Lett. {\bf 125B}, 437 (1983).

\bibitem{Varga1}  K. Varga, R. G. Lovas and R. J. Liotta,
Phys. Rev. Lett. {\bf 69} (1992) 37

\bibitem{Varga2}  K. Varga, R. G. Lovas and R. J. Liotta,
Nucl. Phys. {\bf A550} (1992) 421

\bibitem{talman} J. D. Talman, Nucl. Phys. {\bf A141} (1970) 273

\bibitem{Lawson} R.D. Lawson, Theory of the nuclear shell model,
Clarendon Press, Oxford 1908.

\bibitem{Flies2} R. Blendowske, T. Fliessbach, and H. Walliser,
Z. Phys. A {\bf 339}, 121 (1991).

\bibitem{rose} Rose, Elementary Theory of Angular Momentum,
New York, 1957 J. Wiley and Sons Inc.

\bibitem{vk} The program available on request from the author.
E-mail address: h906var@ella.hu

\end{thebibliography}
\end{document}